\documentclass[aps,floatfix,amsmath,nofootinbib,amssymb,superscriptaddress]{revtex4-2}

\usepackage{overpic}
\usepackage{amssymb}
\usepackage{indentfirst}
\usepackage{feynmf}   %{feynmp}
\usepackage{slashed}  %for Feynman symbols
\usepackage{cases}
\usepackage{color}
\usepackage{multirow}
\usepackage{epstopdf}
\usepackage{graphicx,color,bm}
\usepackage{epstopdf}
\usepackage{booktabs}
\usepackage{float}
\allowdisplaybreaks[4]

\usepackage[colorlinks,
            citecolor=green,
            anchorcolor=red,
            menucolor=red,
            linkcolor=red,
            filecolor=red,
            runcolor=red,
            urlcolor=blue,
            frenchlinks=red]{hyperref}

\begin{document}

\title{T-odd transverse momentum dependent gluon fragmentation functions in a spectator model}

\author{Xiupeng Xie}\affiliation{School of Physics, Southeast University, Nanjing
211189, China}

\author{Zhun Lu}
\email{zhunlu@seu.edu.cn}
\affiliation{School of Physics, Southeast University, Nanjing 211189, China}

\begin{abstract}
We present a model calculation of the T-odd transverse momentum dependent (TMD) gluon fragmentation functions for a spin-1/2 hadron. 
Our model is based on the postulation that a time-like off-shell gluon can fragment into a hadron and a single spectator particle, which is considered to be on-shell. 
We consider the effect of the gluon exchange to calculate all necessary one-loop diagrams for the gluon-gluon correlation functions. 
Two out of four one-loop diagrams give sizeable contributions to the fragmentation functions. 
We obtain analytical expressions for the four T-odd TMD fragmentation functions of the gluon. 
We also provide numerical results on the z-dependence and $k_T$-dependence of the fragmentation functions. 
\end{abstract}

\maketitle

\section{Introduction}

In Quantum Chromodynamics (QCD), the nucleon emerges as a strongly interacting, relativistic bound state of quarks and gluon (collectively called partons). 
This concept of parton dynamics within the framework of QCD replies on two foundamental principles: Color confinement~\cite{tHooft:1973alw,Wilson:1974sk} and asymptotic freedom~\cite{Coleman:1973sx,Gross:2005kv,Gross:1973id,Politzer:1973fx}. 
The former leads to the fact that quarks and gluon are tightly bound within hadrons, rendering the dynamics nonperturbative on the hadronic scale. 
Consequently, the exploration of the partonic structure of hadrons and the mechanisms underlying parton hadronization poses substantial challenges.
Fortunately, the phenomenon of asymptotic freedom, enables the application of QCD factorization to analyze a variety of high-energy scattering processes. 
Within this framework, a collection of well-defined and fundamental functions provide valuable insights into the internal structure of the nucleon and the parton hadronization. These functions are known as the parton distribution functions (PDFs) and the parton fragmentation functions (FFs) respectively.

In general, leading-twist collinear PDFs/FFs are universal between different processes. 
These functions contain the information about 1-dimensional momentum structure. 
A much more comprehensive picture about the partonic structure of hadrons can be achieved by extending the PDFs to the objects in the 3-dimensional momentum space, namely, the transverse momentum dependent (TMD) PDFs and FFs. 
They explicitly depend on the parton transverse momenta and probe the 3-dimensional structure of hadrons~\cite{Collins:2011zzd,Angeles-Martinez:2015sea}. 
Furthermore, TMD PDFs and FFs play important roles in the high energy processes involving two hadrons, such as the semi-inclusive DIS, Drell-Yan process and hadron pair production in $e^+ e^-$ annihilation. 
Different from the collinear PDFs/FFs,  a proper treatment of the process dependent incoming or outgoing directions for the Wilson line~\cite{Collins:2004nx,Collins:2011zzd} of the TMD PDFs or FFs is necessary.

In this paper, we study the T-odd gluon TMD FFs in a spectator model. 
This model is based on the assumption that a time-link off-shell parton can fragment into a hadron and a single real spectator particle. 
This model was originally used to calculate the quark TMD PDFs of the nucleon~\cite{Jakob:1997wg,Brodsky:2002cx,Gamberg:2003ey,Bacchetta:2003rz,
Bacchetta:2008af,Bacchetta:2010si} and has been widely applied in different forms to calculate the quark TMD PDFs of the pion~\cite{Lu:2004hu,Meissner:2008ay,Ma:2019agv}, the TMD FFs of the pion and kaon~\cite{Bacchetta:2007wc}, and the gluon TMD PDFs~\cite{Lu:2016vqu,Bacchetta:2020vty,Bacchetta:2021lvw,Bacchetta:2021twk,Bacchetta:2024fci}. The spectator model has also been extended~\cite{Xie:2022lra} to calculate the T-even gluon TMD FFs $D_1^{h/g} (z,\boldsymbol{k}_T^2)$, $G_{1L}^{h/g} (z,\boldsymbol{k}_T^2)$, $G_{1T}^{h/g} (z,\boldsymbol{k}_T^2)$, and $H_1^{\perp, h/g} (z,\boldsymbol{k}_T^2)$ at the leading-twist level without considering the effects of the gauge-link. 
In this work, in order to obtain the necessary imaginary part required for T-odd gluon TMD FFs $D_{1T}^{\perp, h/g} (z,\boldsymbol{k}_T^2)$, $H_{1T}^{h/g} (z,\boldsymbol{k}_T^2)$, $H_{1L}^{\perp, h/g} (z,\boldsymbol{k}_T^2)$ and $H_{1T}^{\perp, h/g} (z,\boldsymbol{k}_T^2)$, we consider the effect of the gluon rescattering at one loop level. 
The spectator model has also been applied to calculate the Collins function for pions~\cite{Bacchetta:2001di,Bacchetta:2002tk,Gamberg:2003eg,Bacchetta:2003xn,Amrath:2005gv,
Bacchetta:2007wc} and kaons~\cite{Bacchetta:2007wc} by considering the pion loop or the gluon loop.  The twist-3 fragmentation functions $\tilde{H}(z,\boldsymbol{k}_T^2)$, $H(z,\boldsymbol{k}_T^2)$, $\tilde{G}^\perp (z,\boldsymbol{k}_T^2)$ and $G^\perp (z,\boldsymbol{k}_T^2)$~\cite{Lu:2015wja,Yang:2016mxl} have also been calculated within the spectator model.

The rest of the paper is organized as follows. In Section.~\ref{section2}, we provide the formalism of the spectator model with the incorporation the gluon rescattering effect. 
We obtain the model result of the one loop order gluon-gluon correlator and the leading-twist T-odd gluon TMD FFs using proper projecting operators. 
In Section.~\ref{section3}, we present the numerical results of the FFs $D_{1T}^{\perp, h/g} (z,\boldsymbol{k}_T^2)$, $H_{1T}^{h/g} (z,\boldsymbol{k}_T^2)$, $H_{1L}^{\perp, h/g} (z,\boldsymbol{k}_T^2)$ and $H_{1T}^{\perp, h/g} (z,\boldsymbol{k}_T^2)$ in the case the hadron is a proton. 
We summarize the paper in Section.~\ref{section4}.

\section{Analytic calculation of the T-odd FFs of a spin-1/2 hadron}\label{section2}

We follow the conventions in Refs~.\cite{Metz:2016swz,Xie:2022lra} and specify the kinematics of the final-state hadron and the fragmentation parton. 
In a reference frame in which the hadron has no transverse momentum, one can write
\begin{align}
P_{h} & = P_{h}^{-} n_{-}+\frac{M_{h}^{2}}{2 P_{h}^{-}} n_{+}\,,\label{eq:Ph}\\
k & = \frac{P_{h}^{-}}{z} n_{-}+\frac{z\left(k^{2}+\boldsymbol{k}_{T}^{2}\right)}{2 P_{h}^{-}} n_{+}+\boldsymbol{k}_{T}\,,\\
S_{h} & = S_{h L} \frac{P_{h}^{-}}{M_{h}} n_{-}-S_{h L} \frac{M_{h}}{2 P_{h}^{-}} n_{+}+\boldsymbol{S}_{h T}\,,\label{eq:Sh}
\end{align}
where the light-cone vectors $n_{\pm}^2=0$ and $n_+ \cdot n_-=1$ have been used, $M_h$ is the mass of the final-state hadron, $z=P_h^- /k^-$ is the momentum fraction carried by the hadron, $\boldsymbol{k}_T$ denotes the momentum component of the fragmentation parton, $S_{h L}$ and $S_{h T}$ describes longitudinal and transverse polarization of the hadron, respectively. 
The hadron is characterized by its 4-momentum $P_h$ and the covariant spin vector $S_h$.

The appropriate gauge-invariant gluon-gluon correlator for fragmentation is defined as~\cite{Metz:2016swz,Mulders:2000sh}:
\begin{align}
\Delta^{\mu \nu ; \rho \sigma}\left(k ; P_{h}, S_{h}\right) & = \sum_{X} \int \frac{d^{4} \xi}{(2 \pi)^{4}} e^{i k \cdot \xi}\left\langle 0\left|F^{\rho \sigma}(\xi)\right| P_{h}, S_{h} ; X\right\rangle\left\langle P_{h}, S_{h} ; X\left|\mathcal{U}(\xi, 0) F^{\mu \nu}(0)\right| 0\right\rangle\,,
\end{align}
where $\mathcal{U}(\xi, 0)$ is the gauge-link operator connecting space-times $\xi$ and 0 to ensure the gauge-invariance of the operator definition. 
In general, we introduce the correlation function integrated over $k^+$:
\begin{align}
\Delta^{h / g, i j}\left(z, \boldsymbol{k}_{T}^2;S_{h}\right) = \int d k^{+} \Delta^{-j ;-i}\left(k ; P_{h}, S_{h}\right)\,,
\end{align}
where $i$ and $j$ are transverse spatial indices.

We can decompose the gluon fragmentation correlator and define eight leading-twist TMD FFs of the gluon through the following projection~\cite{Collins:1981uw,Collins:2011zzd,Mulders:2000sh}
\begin{align}
\delta_{T}^{i j} \Delta^{h / g, i j}\left(z, \boldsymbol{k}_{T}^2;S_{h}\right)=& 2 P_{h}^{-}\left[D_{1}^{h / g}\left(z, \boldsymbol{k}_{T}^2\right)+\frac{\varepsilon_{T}^{i j} k_{T}^{i} S_{h T}^{j}}{M_{h}} D_{1 T}^{\perp h / g}\left(z, \boldsymbol{k}_{T}^2\right)\right]\,,\label{eq:dD} \\
i \varepsilon_{T}^{i j} \Delta^{h / g, i j}\left(z, \boldsymbol{k}_{T}^2;S_{h}\right)=& 2 P_{h}^{-}\left[\Lambda_{h} G_{1 L}^{h / g}\left(z, \boldsymbol{k}_{T}^2\right)+\frac{\vec{k}_{T} \cdot \vec{S}_{h T}}{M_{h}} G_{1 T}^{h / 77 g}\left(z, \boldsymbol{k}_{T}^2\right)\right]\,, \label{eq:dG} \\
\hat{S} \Delta^{h / g, i j}\left(z, \boldsymbol{k}_{T}^2;S_{h}\right)=& 2 P_{h}^{-} \hat{S}\left[\frac{k_{T}^{i} \varepsilon_{T}^{j k} S_{h T}^{k}}{2 M_{h}} H_{1 T}^{h / g}\left(z, \boldsymbol{k}_{T}^2\right)+\frac{k_{T}^{i} k_{T}^{j}}{2 M_{h}^{2}} H_{1}^{\perp h / g}\left(z, \boldsymbol{k}_{T}^2\right)\right. \notag \\
&\left.+\frac{k_{T}^{i} \varepsilon_{T}^{j k} k_{T}^{k}}{2 M_{h}^{2}}\left(\Lambda_{h} H_{1 L}^{\perp h / g}\left(z, \boldsymbol{k}_{T}^2\right)+\frac{\vec{k}_{T} \cdot \vec{S}_{h T}}{M_{h}} H_{1 T}^{\perp h / g}\left(z, \boldsymbol{k}_{T}^2\right)\right)\right]\,,\label{eq:dH}
\end{align}
and the four T-odd gluon TMD FFs can be projected into
\begin{align}
\frac{\varepsilon^{ij}_{T} k^{i}_{T} S^j_{hT}}{M_h} D_{1T}^{\perp,h/g}\left(z, \boldsymbol{k}_{T}^2\right)=&\frac{1}{2P_h^-} \frac{\delta_{T}^{i j}}{2} \left[\Delta^{h / g, i j}\left(z, \boldsymbol{k}_{T}^2;S_{h}\right)-\Delta^{h / g, i j}\left(z, \boldsymbol{k}_{T}^2;-S_{h}\right)\right]\,,\\
k_{T}^{i}k_{T}^{j}\hat{S}\frac{k_{T}^{i} \varepsilon_{T}^{j k} S_{h T}^{k}}{2 M_{h}} H_{1 T}^{h / g}\left(z, \boldsymbol{k}_{T}^2\right)=&k_{T}^{i}k_{T}^{j}\frac{1}{2P_h^-} \frac{\hat{S}}{2} \left[\Delta^{h / g, i j}\left(z, \boldsymbol{k}_{T}^2;S_{h}\right)-\Delta^{h / g, i j}\left(z, \boldsymbol{k}_{T}^2;-S_{h}\right)\right]\,,\\
S_{T}^{i}S_{T}^{j}\hat{S}\frac{k_{T}^{i} \varepsilon_{T}^{j k} k_{T}^{k}}{2 M_{h}^{2}}\left(\Lambda_{h} H_{1 L}^{\perp h / g}\left(z, \boldsymbol{k}_{T}^2\right)+\right.&\left.\frac{\vec{k}_{T} \cdot \vec{S}_{h T}}{M_{h}} H_{1 T}^{\perp h / g}\left(z, \boldsymbol{k}_{T}^2\right)\right)\notag\\
=&S_{T}^{i}S_{T}^{j}\frac{1}{2P_h^-} \frac{\hat{S}}{2} \left[\Delta^{h / g, i j}\left(z, \boldsymbol{k}_{T}^2;S_{h}\right)-\Delta^{h / g, i j}\left(z, \boldsymbol{k}_{T}^2;-S_{h}\right)\right]\notag\\
-&S_{T}^{i}S_{T}^{j}\hat{S}\frac{k_{T}^{i} \varepsilon_{T}^{j k} S_{h T}^{k}}{2 M_{h}} H_{1 T}^{h / g}\left(z, \boldsymbol{k}_{T}^2\right)\,.
\end{align}
Here, we have used the symmetric transverse tensor $\delta_{T}^{i j} =-g_{T}^{i j}$ with $g_{T}^{i j} = g^{i j}-n_{+}^{i} n_{-}^{j}-n_{+}^{j} n_{-}^{i}$, the anti-symmetric transverse tensor $\epsilon_{T}^{i j}$ with $\epsilon_{T}^{1 2}=1$, and a symmetrization operator $\hat{S}$ for a generic tensor $O^{ij}$ which is defined as:
\begin{align}
\hat{S} O^{i j} \equiv \frac{1}{2}\left(O^{i j}+O^{j i}-\delta_{T}^{i j} O^{k k}\right)\,.
\end{align}

\begin{table}[H]
\centering
\scalebox{1.3}{
\begin{tabular}{|c|c|c|c|}
\hline
$H \backslash g$ & U & Circ & Lin \\
\hline
U & $D_{1}^{h / g}$ &  & $H_{1}^{\perp h / g}$ \\
\hline
L &  & $G_{1 }^{h / g}$ & $H_{1 L}^{\perp h / g}$ \\
\hline
T & $D_{1 T}^{\perp h / g}$ & $G_{1 T}^{h / g}$ & $H_{1}^{ h / g} \quad H_{1 T}^{\perp h / g}$ \\
\hline
\end{tabular}}
\caption{Eight leading-twist TMD FFs of the gluon. The columns indicate the gluon polarization $\textendash$ unpolarized (U), circularly polarized(Circ), linearly polarized (Lin). The rows indicate the hadron prolarization $\textendash$ unpolarized (U), longitudinally polarized (L), transverse polarized (T).}\label{table:func}
\end{table}

\begin{figure}
\centering
\includegraphics[width=0.8\columnwidth]{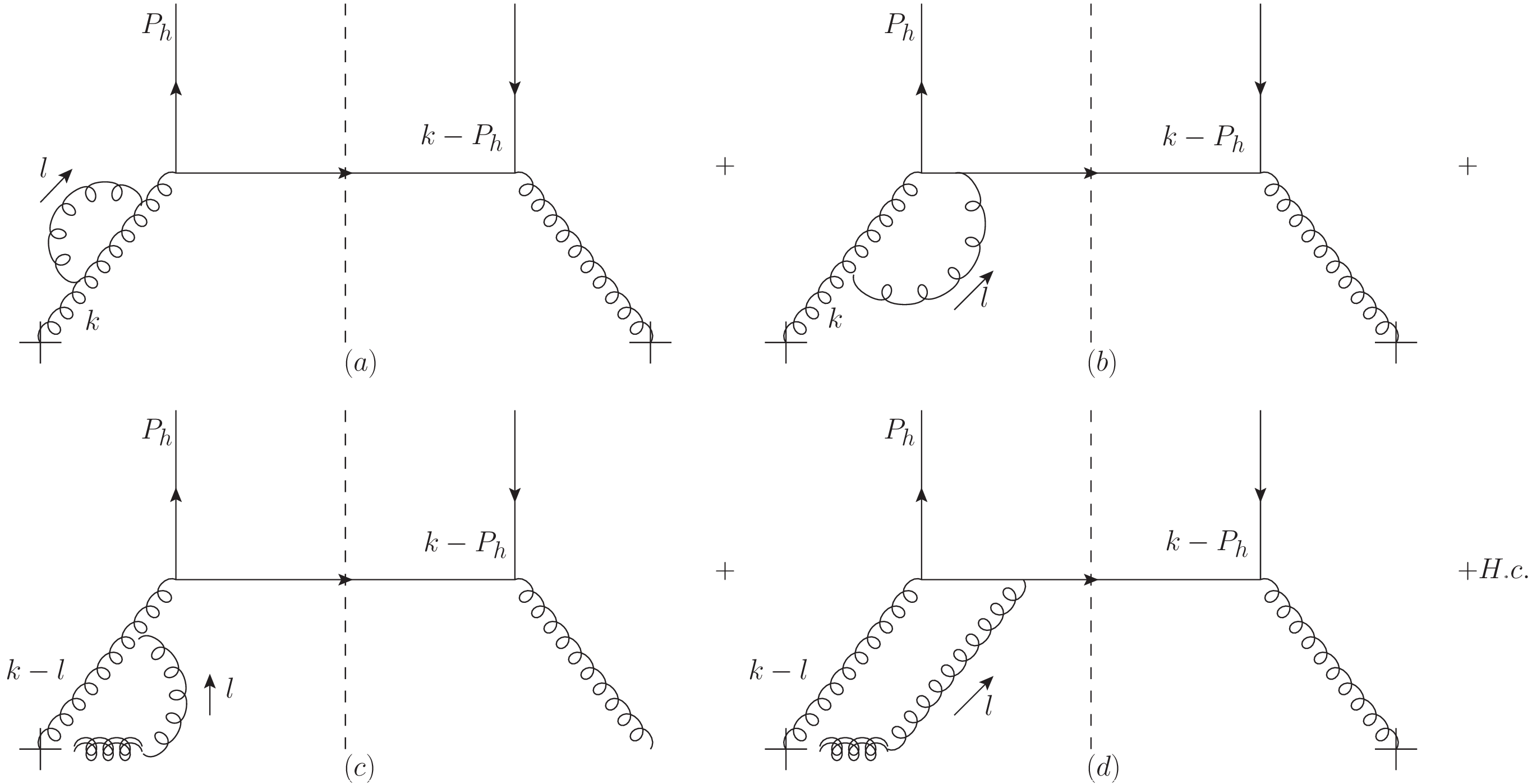}
\caption{One loop order corrections to the fragmentation function of a gluon into a proton in the spectator model. The double gluon lines in (c) and (d) represent the eikonal lines. The double lines correspond to the eikonal lines from the Wilson lines in the definition of gluon-gluon correlator. 
"H.c." stands for the hermitian conjugations of these diagrams.}\label{fig:oneloop}
\end{figure}

The interpretation of the leading-twist TMD FFs in Eqs.~(\ref{eq:dD})-(\ref{eq:dH}) is summarized in Tab.~\ref{table:func}. 
Here $D_1^{h/g}$ is the well-known unpolarized FF which describes the number density of unpolarized hadron in an unpolarized gluon. 
$G_{1L}^{h/g}$, $G_{1T}^{h/g}$, and $H_1^{h/g}$  denote the longitudinally polarized, longi-transversely polarized, and linearly polarized FFs, respectively. 
These functions are T-even, while $D_{1T}^{h/g}$, $H_{1T}^{h/g}$, $H_{1L}^{\perp,h/g}$, $H_{1T}^{\perp,h/g}$ are naive T-odd functions. 
In the spectator model, the tree level gluon-gluon correlator is modeled as:
\begin{align}
\Delta^{i j}\left(z, \boldsymbol{k}_{T}^2; S_h\right) \sim &\frac{1}{(2 \pi)^{3}} \frac{1}{2(1-z) k^{-}}\left[\overline{\mathcal{M}}_0^{j}(z, \boldsymbol{k}_{T}^2;S_h) \mathcal{M}_0^{i}(z, \boldsymbol{k}_{T}^2;S_h)\right]\,.
\end{align}
The tree level diagrams lead to vanishing result for T-odd FFs because of lack of the imaginary phases~\cite{Metz:2002iz,Amrath:2005gv}. 
T-odd functions typically require the interference between two amplitudes with different imaginary parts to exist. 
In order to obtain the necessary imaginary part in the scattering amplitude, one has to consider the diagrams at loop levels. 
In this paper, we will take into account the contribution of gluon rescattering at the one loop order. 
There are four different diagrams (and their hermitian conjugates) that may contribute to the correlator $\Delta^{h / g, i j}\left(z, \boldsymbol{k}_{T}^2;S_{h}\right)$, as shown in Fig.~\ref{fig:oneloop}, including the self-energy diagram (Fig.~\ref{fig:oneloop}a), the vertex diagram (Fig.~\ref{fig:oneloop}b), the hard vertex diagram (Fig.~\ref{fig:oneloop}c), and the box diagram (Fig.~\ref{fig:oneloop}d).

We can write the expressions of the correlator:
\begin{align}
\Delta^{ij}_{(a)} =&\frac{1}{(2 \pi)^3} \frac{1}{2 (1 - z) k^-} \int\frac{d^4 l}{(2 \pi)^4} \operatorname{Tr} \left[ \left( \slashed{P_h} + M_h \right)\frac{1 + \gamma^5 \slashed{S}}{2} \mathcal{Y}_{v, b^{\prime} c^{\prime}}^{*} \frac{G^{j \nu\ast}_{a b^{\prime}} (k, k)}{k^2} \left( \slashed{k} - \slashed{P_h} + M_X \right)_{cc^\prime} \mathcal{Y}_{\mu, b c} \frac{- ig^{\mu \rho} \delta_{bd}}{k^2 + i \varepsilon}\right.\notag \\
&\left.  (-g f_{d^\prime e^\prime f^\prime} V^{\rho^\prime \alpha^\prime \beta^\prime}(k, -l,  l-k)) (-g f_{dfe} V^{\rho \beta \alpha} (-k, k-l, l)) \frac{- ig^{\alpha \alpha^\prime}\delta_{ee^\prime}}{l^2 + i \varepsilon} \cdot \frac{- ig^{\beta \beta^\prime}\delta_{ff^\prime}}{(k - l)^2 + i\varepsilon} \cdot \frac{G^{i \rho^\prime}_{a d^\prime} (k, k)}{k^2} \right] + H.c.\,,\label{eq:delta_a}\\
\Delta^{ij}_{(b)} = & \frac{1}{(2 \pi)^3} \frac{1}{2 (1 - z) k^-} \int\frac{d^4 l}{(2 \pi)^4} \operatorname{Tr} \left[ \left( \slashed{P_h} + M_h \right)\frac{1 + \gamma^5 \slashed{S}}{2} \mathcal{Y}_{v, b^{\prime} c^{\prime}}^{*} \frac{G^{j \nu\ast}_{a b^{\prime}} (k, k)}{k^2} \left( \slashed{k} - \slashed{P_h} + M_X \right)_{c^\prime f^\prime} (g \gamma^{\alpha} f_{f e^\prime f^\prime}) \right.\notag\\
& \left. \frac{i \left( \slashed{k} - \slashed{P_h} - \slashed{l} + M_X \right)_{cf}}{(k- P_h - l)^2 - M_X^2 + i \varepsilon} \mathcal{Y}_{\mu, b c} \frac{-ig^{\rho \rho^\prime} \delta_{bd}}{(k - l)^2 + i \varepsilon} (-g f_{d^\prime ed} V^{\rho^\prime \alpha^\prime \rho} (k, -l, l-k))\frac{- ig^{\alpha \alpha^\prime} \delta_{ee^\prime}}{l^2 + i \varepsilon} \cdot \frac{G^{i \rho^\prime}_{ad^\prime} (k,k)}{k^2} \right] + H.c.\,,\label{eq:delta_b}\\
\Delta^{ij}_{(c)}= & \frac{1}{(2 \pi)^3} \frac{1}{2 (1 - z) k^-} \int\frac{d^4 l}{(2 \pi)^4} \operatorname{Tr} \left[ \left( \slashed{P_h} + M_h \right)\frac{1 + \gamma^5 \slashed{S}}{2} \mathcal{Y}_{v, b^{\prime} c^{\prime}}^{*} \frac{G^{j \nu\ast}_{a b^{\prime}} (k, k)}{k^2} \left( \slashed{k} - \slashed{P_h} + M_X \right)_{c c^\prime} \mathcal{Y}_{\mu, b c} \frac{- ig^{\mu \rho} \delta_{bd}}{k^2 + i \varepsilon} \right.\notag\\
& \left.  (-g f_{d d^\prime f} V^{\rho \rho^\prime \alpha}(-k, k-l, l)) \frac{- ig^{\alpha \rho} \delta_{ff^\prime}}{l^2 + i \varepsilon} \cdot\frac{ign_+^{\rho} f_{eaf^\prime}}{- l \cdot n_+ \pm i \varepsilon} \cdot \frac{G^{i\rho^\prime}_{ad^\prime} (k - l, k - l)}{(k - l)^2 + i \varepsilon} \right] + H.c.\,,\label{eq:delta_c}\\
\Delta^{ij}_{(d)}= & \frac{1}{(2 \pi)^3} \frac{1}{2 (1 - z) k^-} \int\frac{d^4 l}{(2 \pi)^4} \operatorname{Tr} \left[ \left( \slashed{P_h} + M_h \right)\frac{1 + \gamma^5 \slashed{S}}{2} \mathcal{Y}_{v, b^{\prime} c^{\prime}}^{*} \frac{G^{j \nu\ast}_{a b^{\prime}} (k, k)}{k^2} \left( \slashed{k} - \slashed{P_h }+ M_X \right)_{c^\prime f^\prime} \frac{ign_+^{\rho} f_{f d^\prime f^\prime}}{- l \cdot n_+ \pm i \varepsilon}  \right.\notag\\
& \left. \frac{- ig^{\alpha \rho} \delta_{ad^\prime}}{l^2 +i \varepsilon} (g \gamma^{\alpha} f_{f d^\prime f^\prime}) \frac{i \left( \slashed{k} - \slashed{P_h} - \slashed{l} +M_X \right)_{cf}}{(k - P_h - l)^2 - M_X^2 + i \varepsilon} \mathcal{Y}_{\mu, b c} \frac{G^{i \mu}_{ab} (k - l, k - l)}{(k - l)^2 + i \varepsilon}\right] + H.c.\,,\label{eq:delta_d}
\end{align}
where $a,b,c,d,e,f$ are the color indices.
Details of the Feynman rules for the eikonal propagator and the eikonal vertex can be found in Ref.~\cite{Buffing:2017mqm}.
Here, $g$ is the coupling of the three-gluon vertex and the spectator-gluon-spectator vertex, and $M_X$ the mass of the spectator. 
We use the notation
\begin{align}
V^{\mu \nu \rho}(p,q,r)=(p-q)^\rho g^{\mu \nu}+(q-r)^\mu g^{\nu \rho}+(r-p)^\nu g^{\rho \mu}\,.
\end{align}
The same Feynman rules are used here as in Ref.~\cite{Xie:2022lra}. 
The term
\begin{align}
G_{a b}^{i \mu}(k, k) & = -i \delta_{a b}k^{-}\left(g^{i \mu}-\frac{k^{i}n_{+}^{\mu}}{k^{-}}\right)
\end{align}
is the specific Feynman rule for the field strength tensor of the form $-i\left(p^{\mu} g^{\nu \rho}-p^{\nu} g^{\mu \rho}\right) \delta_{a b}$~\cite{Goeke:2006ef,Collins:2011zzd}, and the gluon-hadron-spectator vertex $\mathcal{Y}_{b c}^{\mu}$ is modeled as~\cite{Bacchetta:2020vty}:
\begin{align}
\mathcal{Y}_{b c}^{\mu} & = \delta_{b c}\left[g_{1}\left(k^{2}\right) \gamma^{\mu}+g_{2}\left(k^{2}\right) \frac{i}{2 M_{h}} \sigma^{\mu \nu} k_{\nu}\right]\,,
\end{align}
where $\sigma^{\mu \nu}=i\left[\gamma^{\mu}, \gamma^{\nu}\right]/2$, $g_{1}(k^{2})$ and $g_{2}(k^{2})$ are the gluon-hadron-spectator couplings. 
Here, the vertex is modeled by two Dirac structures $\gamma^{\mu}$ and $\sigma^{\mu \nu} $ which mimics the conserved electromagnetic current of a free nucleon obtained by applying a standard Gordon decomposition. 
The kind of coupling $g_{1,2}(k^{2})$ has several different choices in the literature~\cite{Bacchetta:2008af}. 
Following Refs.~\cite{Bacchetta:2020vty,Xie:2022lra}, we apply the dipolar form factor
\begin{align}
g_{1,2}\left(k^{2}\right) & = \kappa_{1,2} \frac{k^{2}}{\left|k^{2}-\Lambda_{X}^{2}\right|^{2}}\,,
\end{align}
where $\kappa_{1,2}$ are free parameters, and $\Lambda_{X}=k_T^{\textrm{max}}$ is cut-off parameter in order to regularize the divergence. 
The spectator-gluon-spectator vertex is also an important term which connects an anti-octet spectator to an octet gluon and an anti-octet spectator, while the gluon-hadron-spectator vertex connects a color-neutral hadron to an octet gluon and an anti-octet spectator. 
In this study, the spectator-gluon-spectator vertex is modeled as a Dirac structure $g \gamma^{\mu}$ with the color structure constant $f_{abc}$. 
The three-gluon vertex and the $f$-type eikonal vertex of Eqs.~(\ref{eq:delta_c}) and (\ref{eq:delta_d}) will make sense in this case. 

Since the spectator is on-shell $(k-P_h)^2=M_X^2$, we can obtain the following expression for the virtuality $k^2$ of the fragmentation gluon:
\begin{align}
k^{2} = \frac{z}{1-z}\vec{k}_{T}^{2}+\frac{M_X^{2}}{1-z}+\frac{M_{h}^{2}}{z}\,.
\end{align}

In Eqs.~(\ref{eq:delta_c}) and (\ref{eq:delta_d}) we have applied the Feynman rule $i/(- l \cdot n_+ \pm i \varepsilon)$ for the eikonal propagator. It should be noted that the sign of the factor $i \varepsilon$ in the eikonal propagator is different for SIDIS $(+)$ and $e^+ e^-$ annihilation $(-)$. However, in the calculation of T-odd functions, we utilize the Cutkosky cut rule to put certain internal lines on the mass shell to obtain the necessary imaginary phase. 
For the result of Fig.~\ref{fig:oneloop}, the only way of the cuts corresponds to the following replacements on the propagators by using the Dirac delta functions
\begin{align}
\frac{1}{l^2+i \varepsilon} \rightarrow -2\pi i \delta(l^2)\,,\quad \frac{1}{(k-l)^2+i \varepsilon} \rightarrow -2\pi i \delta((k-l)^2)\,.
\end{align}
The other combinations (cutting through the eikonal line or the spectator line) do not contribute, as shown in Refs.~\cite{Yuan:2008it,Lu:2015wja}. 

Now we can obtain the expressions of four the leading-twist T-odd gluon TMD FFs by projecting $\Delta^{ij}$ with $\delta_{T}^{i j}$, $\varepsilon_{T}^{i j} $ and $\hat{S}$ in Eqs.~(\ref{eq:dD})-(\ref{eq:dH}). 
However, using the correlator from Fig.~\ref{fig:oneloop}a or Fig.~\ref{fig:oneloop}c, we obtain the following result
\begin{align}
X^{h/g}\left(z,\boldsymbol{k}^2_T\right) \propto \int \frac{d^4 l}{(2 \pi)^4} \left(k^\alpha l^- - k^- l^\alpha\right) \delta(l^2) \delta((k-l)^2)\,,
\end{align}
Using the decomposition of the integral
\begin{align}
\int d^4 l ~ \delta(l^2) \delta((k-l)^2)l^\mu=\mathcal{F} k^\mu\,,
\end{align}
one can conclude that Eqs.~(\ref{eq:delta_a}) and (\ref{eq:delta_c}) should generate zero contribution to the T-odd TMD FFs. 
Then the result of Eq.~(\ref{eq:delta_b}) reads
\begin{align}
D_{1T(b)}^{\perp,h/g}\left(z, \boldsymbol{k}_{T}^2\right)=&\frac{\left(-2C^2_A C_F\right) g^2 \left(k^2 \left(1-2z\right)+M_h^2-M_X^2\right)}{256 \pi^5 k^4 \left(1-z\right)} D_g\left(z, \boldsymbol{k}_{T}^2\right)\,,\\
H_{1 T(b)}^{h / g}\left(z, \boldsymbol{k}_{T}^2\right)=&-\frac{\left(-2C^2_A C_F\right) g^2 \left(k^2 \left(1-2z\right)+M_h^2-M_X^2\right)}{128 \pi^5 k^4 \left(1-z\right)} D_g\left(z, \boldsymbol{k}_{T}^2\right)\,,\\
H_{1 T(b)}^{\perp h / g}\left(z, \boldsymbol{k}_{T}^2\right)=&\frac{\left(-2C^2_A C_F\right) g^2 M_h^2 z}{64 \pi^5 k^4 \left(1-z\right)} D_g\left(z, \boldsymbol{k}_{T}^2\right)\,,\\
H_{1 L(b)}^{\perp h / g}\left(z, \boldsymbol{k}_{T}^2\right)=&\frac{\left(-2C^2_A C_F\right) g^2 M_h^2}{64 \pi^5 k^4 \left(1-z\right)} D_g\left(z, \boldsymbol{k}_{T}^2\right)\,,
\end{align}
\begin{align}
D_g\left(z, \boldsymbol{k}_{T}^2\right)=&16 g_1^2 M_h^2 \left[\left(\mathcal{B}+\mathcal{Y}\right)+\frac{M_h+M_X}{2M_h}\left(\mathcal{A}+I_2+2\mathcal{Z}\right)\right]-2 g_1 g_2 \left[\left(2\mathcal{A}+\mathcal{B}+4\mathcal{Y}+5\mathcal{Z}\right)k^2\right.\notag\\
&+\left.\left(2\mathcal{A}+5\mathcal{B}+2I_2+4\mathcal{Y}+3\mathcal{Z}\right)\left(M_h^2-M_X^2\right)+\left(4\mathcal{B}-6\mathcal{Y}\right)M_h M_X\right]\notag\\
&-\frac{g_2^2}{M_h}\left[\mathcal{B}\left(k^2\left(M_h-5M_X\right)-\left(M_h-M_X\right)\left(M_h+M_X\right)^2\right)-\mathcal{Y}\left(k^2\left(3M_h-4M_X\right)\right.\right.\notag\\
&+\left.\left.\left(M_h^2-5M_h M_X+4M_X^2\right)\left(M_h+M_X\right)\right)-2\mathcal{Z}k^2\left(2M_h-3M_X\right)\right]\,,
\end{align}
and the result of Eq.~(\ref{eq:delta_d}) has the form 
\begin{align}
&D_{1T(d)}^{\perp,h/g}\left(z, \boldsymbol{k}_{T}^2\right)\notag\\
=&\frac{\left(-2C^2_A C_F\right) g^2 }{128 \pi^5 k^2 \left(1-z\right)z}\left[8g_1^2 M_h \left[z\left(\mathcal{B}M_h\left(z-1\right)-\mathcal{A}M_X\right)+\mathcal{C}k_-\left(z-1\right)\left(M_h\left(z-1\right)+zM_X\right)\right.\right.\notag\\
&+\left.I_2\left(M_h\left(z-1\right)^2+z^2M_X\right)\right]+2g_1 g_2 \left[-z\left(\mathcal{A}k^2\left(2z-1\right)+\mathcal{B}\left(2k^2 \left(z-1\right)+M_h^2\left(2z-1\right)\right.\right.\right.\notag\\
&+\left.\left.\left.3M_h M_X-2M_X^2\left(z-1\right) \right)\right)+\left(\mathcal{C}k_- +I_2\right)\left(z-1\right)\left(k^2-M_h^2\left(2z-1\right)-4M_h M_X-M_X^2\left(z-1\right)\right)\right]\notag\\
&+\frac{g_2^2}{M_h}\left[z\left(\mathcal{A}k^2\left(M_h+3M_X\right)+\mathcal{B}\left(2k^2 M_X+M_h^3+2M_h^2 M_X-M_h M_X^2-2M_X^3\right)\right)\right.\notag\\
&-\left.\left.\left(\mathcal{C}k_- +I_2\right)\left(z-1\right)\left(M_h-M_X\right)\left(k^2-\left(M_h+M_X\right)^2\right)-2I_2 M_X k^2 z\right]\right]\,,\\
&H_{1 T(d)}^{h / g}\left(z, \boldsymbol{k}_{T}^2\right)\notag\\
=&-\frac{\left(-2C^2_A C_F\right) g^2 }{128 \pi^5 k^2 \left(1-z\right)z}\left[16g_1^2 M_h z\left[\mathcal{A}M_h +\mathcal{B}M_h z +\mathcal{C}k_- \left(M_h\left(z-1\right)+zM_X\right)+I_2\left(M_h\left(z-2\right)+zM_X\right)\right]\right.\notag\\
&+4g_1 g_2\left[-z\left(\mathcal{A}\left(k^2\left(2z-1\right)+2\left(M_h^2-M_X^2\right)\right)+\mathcal{B}\left(2z\left(k^2-M_X^2+M_h^2\right)+M_h^2-M_h M_X\right)\right)\right.\notag\\
&+\mathcal{C}k_- \left(\left(1-z\right)\left(k^2+\left(2z+1\right)\left(M_h^2-M_X^2\right)\right)-4zM_h M_X\right)+I_2\left(\left(z+1\right)\left(k^2-\left(2z-1\right)\left(M_h^2-M_X^2\right)\right)\right.\notag\\
&+\left.\left.4z\left(M_h^2-M_X^2-M_h M_X\right)\right)\right]+\frac{2g_2^2}{M_h}\left[z\left(\mathcal{A}k^2\left(3M_h+M_X\right)+\mathcal{B}\left(2k^2+M_h^2-M_X^2\right)\right)\right.\notag\\
&+\left.\left.\left(\mathcal{C}k_- +I_2\right)\left(M_h-M_X\right)\left(k^2 \left(z-1\right)+\left(M_h+M_X\right)\left(M_h\left(z-1\right)+M_X\left(z+1\right)\right)\right)-2I_2 M_h k^2 z\right]\right]\,,\\
&H_{1 T(d)}^{\perp h / g}\left(z, \boldsymbol{k}_{T}^2\right)\notag\\
=&\frac{\left(-2C^2_A C_F\right) g^2 M_h}{32 \pi^5 k^2 \left(1-z\right)^2 z \boldsymbol{k}_T^2}\left[2\left(1-z\right) g_1 g_2 M_h \left[\mathcal{A}\left(z\left(1-z\right)k^2-2z^2\boldsymbol{k}_T^2 \right)+z\boldsymbol{k}_T^2\left(\left(1-z\right)\mathcal{C}k_-+\left(1+z\right)I_2\right)\right.\right.\notag\\
&+\left.\mathcal{B}M_h^2 \left(1-z\right)\right]+g_2^2\left[\mathcal{A}M_X z \left(1-z\right)k^2+\mathcal{B}M_h\left(1-z\right)\left(M_h M_X-z^2 \boldsymbol{k}_T^2\right)\right.\notag\\
&+\left.\left.z\boldsymbol{k}_T^2 \left(\mathcal{C}k_-+I_2\right)\left(M_h \left(1-z\right)^2+M_X\left(1-z^2\right)\right)\right]\right]\,,\\
&H_{1 L(d)}^{\perp h / g}\left(z, \boldsymbol{k}_{T}^2\right)\notag\\
=&\frac{\left(-2C^2_A C_F\right)g^2 g_1^2 M_h^2 z \left(\mathcal{A}-I_2\right)}{8 \pi^5 \left(1-z\right) k^2}-\frac{\left(-2C^2_A C_F\right)g^2 }{128 \pi^5 z^2 \left(1-z\right) k^2 \boldsymbol{k}_T^2}\left[2g_1 g_2 M_h\left[2\left(M_h\left(z-1\right)+M_X z\right)\left(\mathcal{A}z k^2 +\mathcal{B} M_h^2\right)\right.\right.\notag\\
&+\left.z\boldsymbol{k}_T^2\left(2z M_h \left(4\mathcal{A}+\mathcal{B}\right)+\mathcal{C}k_-\left(5M_h\left(z-1\right)+8M_X z^2\right)+\mathcal{D}k_- M_h \left(z-1\right)z+I_2\left(8M_X z^2-M_h\left(3z+5\right)\right)\right)\right]\notag\\
&-g_2^2\left[2\left(M_h-M_X\right)\left(M_h\left(z-1\right)+zM_X\right)\left(\mathcal{A}zk^2+\mathcal{B}M_h^2+2z\left(z-1\right)\boldsymbol{k}_{T}^2\left(\mathcal{C}k_- +I_2\right)\right)\right.\notag\\
&+z\boldsymbol{k}_{T}^2\left(4zk^2\left(z\left(\mathcal{A}+\mathcal{B}\right)-I_2\right)-4\mathcal{C}k_- z \left(1-z\right)k^2+\left(\mathcal{C}k_-+I_2\right)M_h\left(9M_X-M_h\right)-2\mathcal{B}z\left(M_h^2\left(3-2z\right)\right.\right.\notag\\
&+\left.\left.\left.\left.M_h M_X+2z M_X^2\right)-\mathcal{D}k_- z M_h \left(M_h-M_X\right)\right)\right]\right]\,.
\end{align}
Here the functions $I_i$ are defined as
\begin{align}
I_1=&\int d^4 l ~ \delta(l^2) \delta((k-l)^2-m^2)=\frac{\pi}{2k^2}\left(k^2-m^2\right)\,,\\
I_2=&\int d^4 l ~\frac{\delta(l^2) \delta((k-l)^2-m^2)}{\left(k-P_h-l\right)^2-M_X^2}=\frac{\pi}{2\sqrt{\lambda \left(M_h,M_X\right)}} \ln\left(1-\frac{2\sqrt{\lambda \left(M_h,M_X\right)}}{k^2-M_h^2+M_X^2+\sqrt{\lambda \left(M_h,M_X\right)}}\right)\,,\\
I_3=&\int d^4 l ~\frac{\delta(l^2) \delta((k-l)^2-m^2)}{- l \cdot n_+ + i \varepsilon}\,,\\
I_4=&\int d^4 l ~\frac{\delta(l^2) \delta((k-l)^2-m^2)}{\left(- l \cdot n_+ + i \varepsilon\right)\left(\left(k-P_h-l\right)^2-M_X^2\right)}\,,
\end{align}
with $\lambda \left(M_h,M_X\right)=\left(k^2-\left(M_h+M_X\right)^2\right)\left(k^2-\left(M_h-M_X\right)^2\right)$, and $I_{34}$ is the linear combination of $I_3$ and $I_4$,
\begin{align}
I_{34}=k_-\left(I_3+(1-z)(k^2-m^2)I_4\right)=\pi \ln\frac{\sqrt{k^2}(1-z)}{M_X}\,.
\end{align}

$\mathcal{A}$ and $\mathcal{B}$ denote the following functions
\begin{align}
\mathcal{A}=&\frac{I_1}{\lambda \left(M_h,M_X\right)}\left(2k^2\left(k^2-M_h^2-M_X^2\right)\frac{I_2}{\pi}+\left(k^2+M_h^2-M_X^2\right)\right)\,,\\
\mathcal{B}=&-\frac{2k^2}{\lambda \left(M_h,M_X\right)} I_1 \left(1+\frac{k^2-M_h^2+M_X^2}{\pi}I_2\right)\,,
\end{align}
which appear in the decomposition of the integral
\begin{align}
\int d^4 l ~\frac{l^\mu \delta(l^2) \delta((k-l)^2-m^2)}{\left(k-P_h-l\right)^2-M_X^2}=\mathcal{A}k^\mu +\mathcal{B} P_h^\mu\,.
\end{align}

$\mathcal{C}$, $\mathcal{D}$ and $\mathcal{E}$ denote the following functions
\begin{align}
\mathcal{C}P_h^-=&\frac{I_{34}}{2\boldsymbol{k}_T^2}+\frac{1}{2z\boldsymbol{k}_T^2}\left[-zk^2+(2-z)M_h^2+zM_X^2\right]I_2\,,\\
\mathcal{D}P_h^-=&\frac{-I_{34}}{2z\boldsymbol{k}_T^2}-\frac{1}{2z\boldsymbol{k}_T^2}\left[(1-2z)k^2+M_h^2-M_X^2\right]I_2\,,\\
\mathcal{E}k^-=&\frac{\lambda \left(M_h,M_X\right)}{4zP_h^- \boldsymbol{k}_T^2}I_2-\frac{1}{4z^2\boldsymbol{k}_T^2}\left[(1-2z)k^2+M_h^2-M_X^2\right]I_{34}+\frac{k^2-m^2}{2}I_4\,,
\end{align}
which appear in the decomposition of the integral
\begin{align}
\int d^4 l ~\frac{l^\mu \delta(l^2) \delta((k-l)^2-m^2)}{\left(- l \cdot n_+ + i \varepsilon\right)\left(\left(k-P_h-l\right)^2-M_X^2\right)}=\mathcal{C}k^\mu+\mathcal{D}P_h^\mu+\mathcal{E} n_+^\mu\,.
\end{align}
$\mathcal{X}$, $\mathcal{Y}$ and $\mathcal{Z}$ denote the following functions
\begin{align}
\mathcal{X}=&\frac{I_1^2}{\pi M_X^2\lambda \left(M_h,M_X\right) }\left[(k^2+M_X^2)(M_h^2+2M_X^2)-3M_X^4-k^4\right]\,,\\
\mathcal{Y}=&\frac{I_1^2}{\pi M_X^2\lambda \left(M_h,M_X\right) }\left[k^2(M_h^2-M_X^2)-k^4\right]\,,\\
\mathcal{Z}=&\frac{I_1^2}{\pi M_X^2\lambda \left(M_h,M_X\right) }\left[k^4-k^2(M_h^2+M_X^2)\right]\,,
\end{align}
which appear in the decomposition of the integral
\begin{align}
\int d^4 l ~\frac{l^\mu l^\nu \delta(l^2) \delta((k-l)^2-m^2)}{\left(k-P_h-l\right)^2-M_X^2}=\mathcal{X}k^\mu k^\nu +\mathcal{Y} P_h^\mu P_h^\nu+\mathcal{Z}\left(k^\mu P_h^\nu+P_h^\mu k^\nu\right)\,.
\end{align}

\section{Numerical Results}\label{section3}
\begin{table}[H]
	\centering
	\renewcommand\tabcolsep{15.0pt}
	\renewcommand{\arraystretch}{2}
	\begin{tabular}{ccccc}
		\toprule[2pt]
		$\kappa_{1}^{p}$&   $\kappa_{2}^{p}$ & $\Lambda_{X}^{p}$& $M_X^{p}$ &$\chi^2/\mathrm{d.o.f.}$\\
		\midrule[1pt]
        7.742$\pm$0.460&2.238$\pm$1.563 & 1.589$\pm$0.014 &1.252$\pm$0.016 & 2.906\\
		\bottomrule[2pt]
	\end{tabular}
	\caption{Fitted values of the parameters in the spectator model using the AKK08 parametrization for the gluon FFs.}
	\label{table:para}
\end{table}

In this section, we present the numerical results of the T-odd gluon fragmentation functions $D_{1T}^{\perp, h/g} (z,\boldsymbol{k}_T^2)$, $H_{1T}^{h/g} (z,\boldsymbol{k}_T^2)$, $H_{1L}^{\perp, h/g} (z,\boldsymbol{k}_T^2)$ and $H_{1T}^{\perp, h/g} (z,\boldsymbol{k}_T^2)$. 
To do this the integrated FFs are defined as
\begin{align}
D_{1T}^{\perp(1), h/g} (z)=&z^2 \int d^2 \boldsymbol{k}_T \frac{\boldsymbol{k}_T^2}{2M_h^2} D_{1T}^{\perp, h/g} (z,\boldsymbol{k}_T^2)\,,\\
H_{1T}^{h/g} (z)=&z^2 \int d^2 \boldsymbol{k}_T  H_{1T}^{h/g} (z,\boldsymbol{k}_T^2)\,,\\
H_{1L}^{\perp(1), h/g} (z)=&z^2 \int d^2 \boldsymbol{k}_T \frac{\boldsymbol{k}_T^2}{2M_h^2} H_{1L}^{\perp, h/g} (z,\boldsymbol{k}_T^2)\,,\\
H_{1T}^{\perp, h/g} (z)=&z^2 \int d^2 \boldsymbol{k}_T H_{1T}^{\perp, h/g} (z,\boldsymbol{k}_T^2)\,.
\end{align}
Note that in the reference of Eqs.~(\ref{eq:Ph})-(\ref{eq:Sh}) the fragmentation parton has the transverse momentum $\boldsymbol{k}_T$ and the hadron has no transverse momentum, while in a frame of reference in which the fragmentation parton has no transverse momentum and the hadron has the transverse momentum $\boldsymbol{P}_{hT}$, one can show that $\boldsymbol{P}_{hT}=-z\boldsymbol{k}_T$, and $z^2 \boldsymbol{k}_T=\boldsymbol{P}_{hT}^2$.

In this work, we take proton as the spin-$1/2$ hadron ($h\equiv p$) to perform the calculation. The original values of the parameters (AKK08) from Ref.~\cite{Xie:2022lra} are presented in Table~\ref{table:para}. 
In the fit we adopt the AKK08 parametrization at the scale $Q_0=1.5~\mathrm{GeV}$ to avoid negative values for $D_1^{h/g}(z)$ in the small $z$ region and low-$Q$. 
And we consider the range $0.1<z<0.7$ in the fit. 
In the following, the strong coupling constant is fixed to $\alpha(Q_0)=g^2/4\pi=0.35$.

\begin{figure}
    \centering
    \includegraphics[width=0.45\columnwidth]{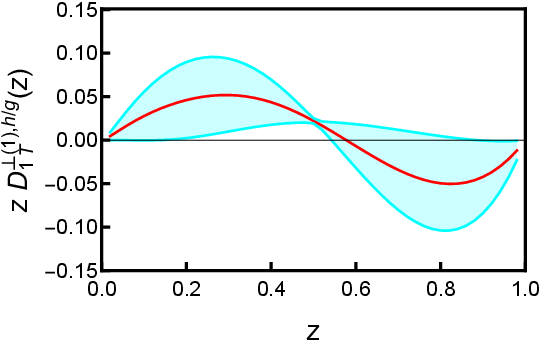}
    \quad
    \includegraphics[width=0.45\columnwidth]{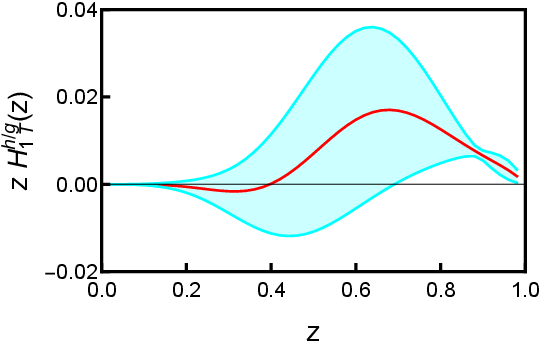}
    \quad
    \includegraphics[width=0.45\columnwidth]{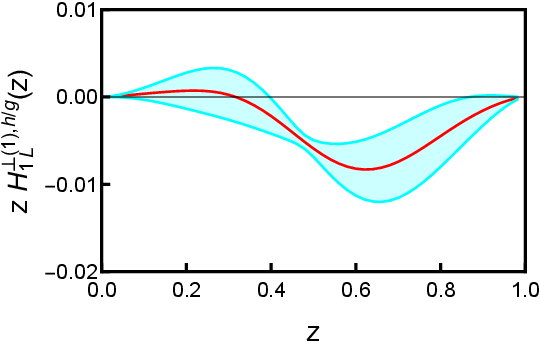}
    \quad
    \includegraphics[width=0.45\columnwidth]{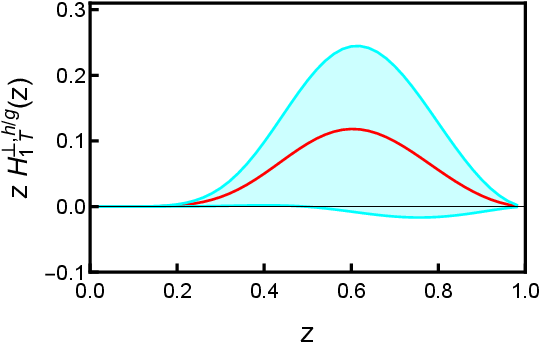}
    \caption{T-odd gluon fragmentation functions $zD_{1T}^{\perp(1), h/g} (z)$ (upper-left), $zH_{1T}^{h/g} (z)$ (upper-right), $zH_{1L}^{\perp(1), h/g} (z)$ (lower-left), and $zH_{1T}^{\perp, h/g} (z)$ (lower-right) vs $z$, respectively, at $Q_0=1.5~\mathrm{GeV}$ in the spectator model. The solid lines depict the spectator model results. The bands depict the uncertainties from the uncertainties of the parameters.}
    \label{fig:zfunc}
\end{figure}

In Fig.~\ref{fig:zfunc} we present the numerical results of $zD_{1T}^{\perp(1), h/g} (z)$, $zH_{1T}^{h/g} (z)$, $zH_{1L}^{\perp(1), h/g} (z)$, and $zH_{1T}^{\perp, h/g} (z)$ vs $z$ using the parameters in Table~\ref{table:para}. 
We find that the magnitudes of $zD_{1T}^{\perp(1), h/g} (z)$ and $zH_{1T}^{\perp, h/g} (z)$ are sizable. 
The sizes of $zH_{1T}^{h/g} (z)$ and $zH_{1L}^{\perp(1), h/g} (z)$ are several times less than that of $zD_{1T}^{\perp(1), h/g} (z)$ and $zH_{1T}^{\perp, h/g} (z)$. 
This finding indicates that the effects of these fragmentation functions could be significant and can be probed in future experimental measurements. 
There is a node at $z=0.6$ for $zD_{1T}^{\perp(1), h/g} (z)$. 
The sign of $zD_{1T}^{\perp(1), h/g} (z)$ flip when $z$ increases from lower region to higher region, while $zH_{1T}^{\perp, h/g} (z)$ is positive in the entire $z$ region. 
Moreover, we observe that $zH_{1T}^{h/g} (z)$ and $zH_{1L}^{\perp(1), h/g} (z)$ have the opposite trend, while their signs flip at $0.3<z<0.4$.

\begin{figure}
    \centering
    \includegraphics[width=0.45\columnwidth]{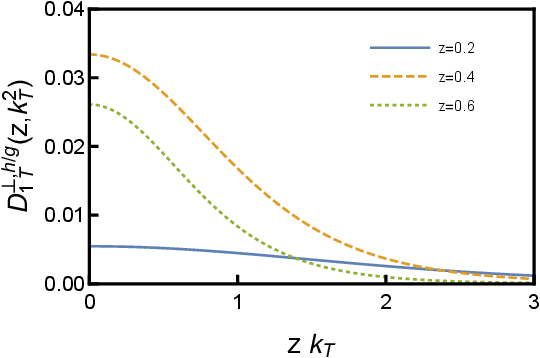}
    \quad
    \includegraphics[width=0.45\columnwidth]{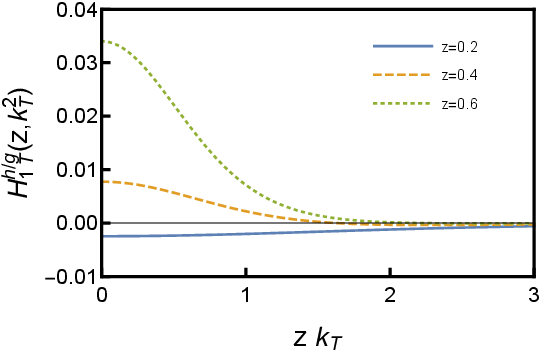}
    \quad
    \includegraphics[width=0.45\columnwidth]{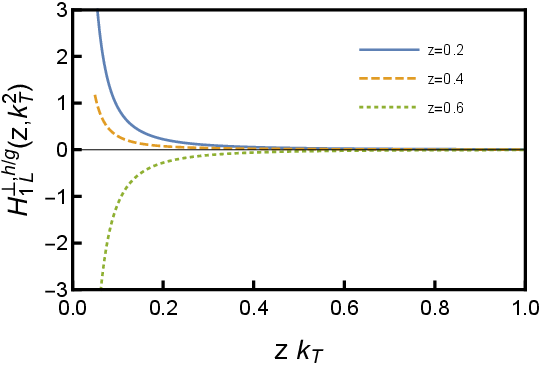}
    \quad
    \includegraphics[width=0.45\columnwidth]{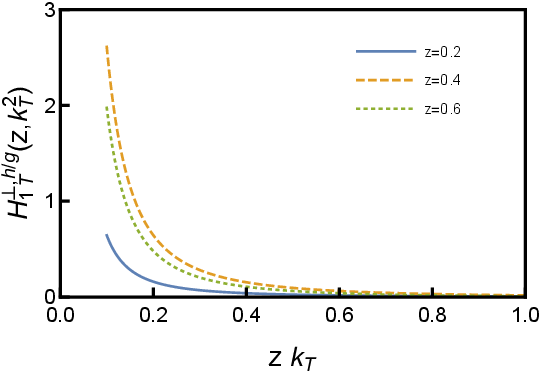}
    \caption{T-odd gluon fragmentation functions $D_{1T}^{\perp, h/g} (z,\boldsymbol{k}_T^2)$ (upper-left), $H_{1T}^{h/g} (z,\boldsymbol{k}_T^2)$ (upper-right), $H_{1L}^{\perp, h/g} (z,\boldsymbol{k}_T^2)$ (lower-left), and $H_{1T}^{\perp, h/g} (z,\boldsymbol{k}_T^2)$ (lower-right) vs $z\boldsymbol{k}_T$  in the case $h\equiv p$ at $z=0.2$, $0.4$ and $0.6$, respectively.}
    \label{fig:ktfunc}
\end{figure}

In Fig.~\ref{fig:ktfunc}, we depict the transverse momentum dependence of the four T-odd TMD FFs of the gluon as functions of $z\boldsymbol{k}_T=\left|\boldsymbol{P}_{hT}\right|$ at $z=0.2$, $0.4$ and $0.6$, respectively. 
These functions have quite different sizes and TMD-shapes at different $z$ values. Furthermore, the sizes of $D_{1T}^{\perp, h/g} (z,\boldsymbol{k}_T^2)$ and $H_{1T}^{h/g} (z,\boldsymbol{k}_T^2)$ decrease smoothly with increasing $z\boldsymbol{k}_T$, while 
$H_{1L}^{\perp, h/g} (z,\boldsymbol{k}_T^2)$ and $H_{1T}^{\perp, h/g} (z,\boldsymbol{k}_T^2)$ have larger values in the small $z\boldsymbol{k}_T$ region and decrease rapidly in the large $z\boldsymbol{k}_T$ region.

For fragmentation functions, positivity bounds are important model-independent constraints. For gluon TMD FFs, the constraints become~\cite{Mulders:2000sh,Bacchetta:1999kz}
\begin{align}
\frac{\boldsymbol{k}_T^2}{M_h^2}\left(\left|G_{1T}^{\perp,h/g}(z,\boldsymbol{k}_T^2)\right|^2+\left|D_{1T}^{\perp, h/g} (z,\boldsymbol{k}_T^2)\right|^2\right) &\le D_{1}^{h/g}(z,\boldsymbol{k}_T^2)\,,\label{eq:bound1}\\
\frac{\boldsymbol{k}_T^4}{4 M_h^4}\left(\left|H_{1}^{\perp,h/g}(z,\boldsymbol{k}_T^2)\right|^2+\left|H_{1L}^{\perp, h/g} (z,\boldsymbol{k}_T^2)\right|^2\right) &\le D_{1}^{h/g}(z,\boldsymbol{k}_T^2)\,,\label{eq:bound2}\\
\frac{\boldsymbol{k}_T^4}{4 M_h^4} \left| H_{1T}^{\perp, h/g} (z,\boldsymbol{k}_T^2)\right| &\le \frac{\left|\boldsymbol{k}_T\right|}{2 M_h} \left(D_{1}^{h/g}(z,\boldsymbol{k}_T^2)-G_{1L}^{\perp,h/g}(z,\boldsymbol{k}_T^2)\right)\,,\label{eq:bound3}\\
\frac{\boldsymbol{k}_T^2}{2 M_h^2} \left|H_{1T}^{h/g} (z,\boldsymbol{k}_T^2)+ \frac{\boldsymbol{k}_T^2}{2 M_h^2} H_{1T}^{\perp, h/g} (z,\boldsymbol{k}_T^2)\right|&\le \frac{\left|\boldsymbol{k}_T\right|}{2 M_h} \left(D_{1}^{h/g}(z,\boldsymbol{k}_T^2)+G_{1L}^{\perp,h/g}(z,\boldsymbol{k}_T^2)\right)\,,\label{eq:bound4}
\end{align}
We have conducted numerical verifications and found that our model results for the T-odd gluon TMD FFs $D_{1T}^{\perp, h/g} (z,\boldsymbol{k}_T^2)$, $H_{1T}^{h/g} (z,\boldsymbol{k}_T^2)$, $H_{1L}^{\perp, h/g} (z,\boldsymbol{k}_T^2)$, and $H_{1T}^{\perp, h/g} (z,\boldsymbol{k}_T^2)$ satisfy the bounds in Eqs.~(\ref{eq:bound1})-(\ref{eq:bound4}).

\section{Conclusions}\label{section4}

In this work, we have studied the T-odd gluon fragmentation functions $D_{1T}^{\perp, h/g} (z,\boldsymbol{k}_T^2)$, $H_{1T}^{h/g} (z,\boldsymbol{k}_T^2)$, $H_{1L}^{\perp, h/g} (z,\boldsymbol{k}_T^2)$, and $H_{1T}^{\perp, h/g} (z,\boldsymbol{k}_T^2)$ in a spectator model, which is based on a assumption that a time-like off-shell gluon can fragment into a hadron and a single spectator particle. 
The tree level diagrams lead to a vanishing result of T-odd FFs because of lack of the imaginary phase. 
We considered the effect gluon exchange to calculate all necessary one-loop diagrams for the gluon-gluon correlation functions. 
We found that two of them have nonzero contribution to T-odd gluon TMD FFs.
We obtained the analytical expressions of four T-odd gluon TMD FFs by projecting the correlators to the symmetric and antisymmetric tensors $\delta_T^{ij}$, $\epsilon_T^{ij}$ and $\hat S$. 
In the calculation we adopted a dipolar form factor for the gluon-hadron-spectator coupling and a  spectator-gluon-spectator vertex. 
With the parameters fitted from the AKK08 parametrization, we presented the numerical results of the $z$-dependence and TMD-dependence of the FFs $D_{1T}^{\perp, p/g}$, $H_{1T}^{p/g}$, $H_{1L}^{\perp, p/g}$, and $H_{1T}^{\perp, p/g}$. 
We also checked the positivity bounds for T-odd FFs showing that our model results satisfy these bounds. 
Our investigation indicates that the magnitudes of the T-odd gluon FFs could be significant, and can provides useful information for future experimental measurements and theoretical model improvements.

\section*{Acknowledgements}
This work is partially supported by the National Natural Science Foundation of China under grant number 12150013.

\end{document}